\documentstyle[11pt,newpasp,twoside,epsf]{article}
\def\eg{e.g., }
\def\etal{et~al.\ }
\def\edcomment#1{\iffalse\marginpar{\raggedright\sl#1\/}\else\relax\fi}
\reversemarginpar

\begin{document}
\title{Intracluster Planetary Nebulae in Clusters and Groups}
\author{John J. Feldmeier}
\affil{Department of Astronomy, Case Western Reserve University, 
Cleveland, OH, USA}

\author{Robin B. Ciardullo}
\affil{Department of Astronomy and Astrophysics, Pennsylvania State
University, University Park, PA}

\author{George H. Jacoby}
\affil{WIYN Observatory, Tucson, AZ, USA}

\author{Patrick R. Durrell}
\affil{Department of Astronomy and Astrophysics, Pennsylvania State
University, University Park, PA}

\author{J. Christopher Mihos}
\affil{Department of Astronomy, Case Western Reserve University, 
Cleveland, OH, USA}

\begin{abstract}
We present the results from multiple surveys for intracluster planetary
nebulae (IPNe) in nearby galaxy clusters and groups.  We find that
in the case of clusters, our observations imply: 1) the amount of
intracluster starlight is significant, up to 20\% of the total starlight,
2) the Virgo Cluster is elongated along our line of sight, and 3) 
the intracluster light is clustered
on the sky, implying ongoing tidal stripping.  In contrast, searches
for IPNe in groups have found little or no intra-group population, implying
there may be something in the cluster environment that significantly
enhances intracluster star production.  From high-resolution N-body 
simulations, we find that the IPNe should create observable features 
in position-velocity space, and that these features may eventually
allow us to place limits on the dynamics of galaxy clusters.

\end{abstract}

\section{Introduction}

The study of intracluster starlight (ICL) has grown dramatically
in the last few years.  Once thought to be just another 
odd prediction of Zwicky (1951), intracluster starlight may be a 
useful tool in understanding the evolution of galaxies in clusters, 
and may be an important chain in the recycling of intergalactic
and interstellar matter (see Arnaboldi, this conference for a review).

In particular, intracluster planetary nebulae (IPNe) are an 
excellent tracer of the intracluster light, and can be detected
relatively easily in nearby galaxy clusters with deep narrow-band
imaging.  Extragalactic planetaries appear as point sources through
a [O~III] $\lambda$5007 narrow-band filter, but disappear altogether
when imaged through a ``off-band'' filter.  
Planetary nebulae also follow the [O~III] $\lambda$5007 planetary
nebulae luminosity function (PNLF), which is a highly accurate
distance indicator (see Ciardullo \etal 2002b and references therein).  
The well-defined luminosity function allows us to gather depth information 
on the intracluster stars.  Finally, since IPNe are emission-line objects, 
their radial velocities can be determined with 
moderate-resolution spectroscopy (Freeman \etal 2000), 
allowing us to obtain crucial dynamical information.

Here, we focus on our group's efforts to search for IPNe in 
nearby galaxy clusters and groups, and give a brief summary on the
results to date.  It is important to stress that surveys
for IPNe are not pristine: we estimate that about 20\% of our
IPNe candidates in Virgo are actually Lyman-$\alpha$ galaxies at
z = 3.1 (Ciardullo \etal 2002a).    

\section{Studies in Clusters}

IPNe were first discovered in the Virgo cluster as a population
of ``overluminous'' planetary nebulae, 
though they were not originally recognized 
as such (Jacoby, Ciardullo, \& Ford 1990).
Kinematic proof of IPNe was then found by Arnaboldi \etal (1996),
and IPNe were subsequently detected in the 
Fornax cluster by Theuns \& Warren (1997).  Additional
evidence for large numbers of IPNe in front of the Virgo elliptical M~87
quickly followed (Ciardullo \etal 1998).  With the advent of wide-field
mosaic detectors, it became feasible to observe much larger portions of
the Virgo and Fornax clusters for IPNe.  

Figure~1 shows the status of our surveys to date.  We have detected 318
IPNe candidates in the Virgo cluster (Feldmeier, Ciardullo, \& Jacoby 1998;
Feldmeier \etal 2003; Feldmeier \etal 2004), and 95 candidates in the
Fornax cluster (Ciardullo \etal 2004).  Transforming the numbers of
IPNe to a stellar luminosity is complicated by several factors: the
amount of background contamination, the known density differences of
PNe to stellar luminosity (Ciardullo 1995), and any line-of-sight 
effects (Feldmeier \etal 2004).  However,
if we take conservative limits for such effects, we find that both 
clusters contain 10--20\% of their stars in an intracluster component.  

The spatial distribution of IPNe in the cluster is also of great interest,
since PNe closely follow the starlight in galaxies (Ciardullo \etal 1989). 
For instance, in Virgo we can find the upper limit distance to 
each IPNe field using the sharp cutoff of the PNLF.  We can then
compare these distances to the PNLF distances of the cluster ellipticals
(Jacoby, Ciardullo, \& Ford 1990; Ciardullo \etal 1998) and to the
HST Cepheid distances of spirals (Freedman \etal 2001).  As demonstrated
in Figure 2, the IPNe are enormously extended, up to 3 Mpc in depth.
This agrees with the inferred depth of Virgo's spiral galaxies (derived
from the Tully-Fisher observations; Solanes \etal 2002), and suggests
that the bulk of the IPNe come from late-type galaxies

When we compare the IPNe density in Virgo and Fornax 
to that of the galaxies directly, 
we find that in both clusters, the amount of IPNe may drop more 
slowly with radius than that of the galaxies, though the 
scatter is large due to small numbers.  From observing the 
positions of the IPNe on the sky, we also gain some
insight on the spatial distribution of the ICL.  We find that the IPNe
are clustered on arcminute scales, implying that there is ongoing
tidal stripping in these clusters.

\section{Studies in Groups}

Although the amount of intracluster starlight in clusters such as
Virgo and Fornax is now well established, the amount of 'intra-group'
starlight is still uncertain.  Theoretical studies predict that 
if most intracluster stars are removed by galaxy collisions 
(\eg Richstone \& Malumuth 1983; Moore \etal 1996), the 
fraction of intra-group stars, to first order, should be a 
smooth function of galaxy number density (L$_{{ICL}}$ $\sim$ 
N$_{{Gal}}^{2}$).

To test this hypothesis, we have undertaken a large-scale 
[O~III] $\lambda$5007 IPN survey of the nearby M~81 group of galaxies.  
This galaxy group is known to be strongly interacting, with 
multiple tidal tails seen in H~I gas (Yun, Ho, \& Lo 1994).  We have
surveyed 1.44 square degrees of this system with the KPNO 4-m and the
Mosaic camera, and have reached at least two magnitudes down the PNLF
in all of our fields.  

Although the analysis is ongoing, there is already a clear result: there
is substantially less IPNe in the M~81 group than in the rich clusters.  
For example, if we take the results from our Field~1, we find 102 PNe 
candidates near M81 proper, but no objects whatsoever in the 
remaining half of the field.  If we assume this density limit is
typical, and adopt a limit of 1 $\pm$ 1 intra-group PNe 
this leads us to a intra-group fraction of 1.3\%.  These
results are strengthened by a deep broad-band survey of the M81
group to look for intra-group red giant stars which find a similarly
small limit (see Durrell, this volume).

Our result, combined with a similar result for the Leo~I Group (1.6\%; 
Castro-Rodr{\'{\i}}guez \etal 2003), strongly implies there is substantially
less (4--15 times) less intergalactic stars in groups than there
are in clusters.  The drop in density is unexpected, and will need
to be explained by models of intracluster star production (see
Ciardullo, this volume).

One of the motivations for surveying the M81 and Leo I systems for
intra-group PNe is
that both these groups contain a network of H I tidal features
(Schneider \etal 1989; Yun, Ho, \& Lo 1994).  However, no
correlation has been found between the H I gas and the IPNe.  It may be
that for normal spiral galaxies, tidal interactions can remove the
H~I in the outer galaxy without causing any of the stars in the
inner regions to escape.  IPNe observations in at least one
undisturbed galaxy group would be helpful to ensure this result.

\section{Dynamical Studies of Intracluster Light}

Large samples of IPNe are already available for a radial
velocity measurements, and many more will be found in the near
future.  Consequently, it is important to understand the
kinematics of the ICL.  Several groups have already
begun the process of modeling the velocity structure of the
ICL through a variety
of simulations (Moore \etal 1996; Dubinski, Murali, \& Ouyed 2001; 
Napolitano \etal 2003).  In these 
models, the intracluster light generally follows a radial orbit 
distribution, but is dynamically unrelaxed, and fills the observed phase 
space non-uniformly, due to the presence of tidal debris.  The radial
orbit envelope may allow us to estimate the mass profile
of galaxy clusters (Dubinski, Murali, \& Ouyed 2001), independent 
of more traditional methods.  Moreover, since the phase-space clumpiness 
of the ICL is related to the cluster's dynamical age, the simulations, 
combined with the IPN observations,  will allow us to place interesting 
limits on this quantity, as well as potential limits on cosmological 
parameters.

To improve upon the situation, we have constructed our own
high-resolution, fully-self consistent N-body models, using large
scale structure simulations.  We use the simulations to find
proto-clusters at high redshift, replace the low resolution
galaxies with self-consistent galactic models (Hernquist 1993),
and run the simulations.  We confirm the general
results thus far: our ongoing focus is on developing useful metrics for
observations.

\begin{figure}
%\plotfiddle{ipn.ps}{0pt}{0}{46}{46}{-225}{-325}
%\plotfiddle{fornax_vert.ps}{0pt}{0}{45}{45}{-40}{-290}
\vspace{270pt} 
\protect\caption{
Regions of the Virgo (left) and Fornax (right) clusters, drawn
from the Digital Sky Survey, with the location of our IPN survey 
fields marked.  Arrows represent additional detections of ICL made 
by other researchers.  Thus far, we have found a total of 
413 IPN candidates in both clusters, but we have observed less than
1\% of each cluster.}
\end{figure}

\begin{figure}
\plotone{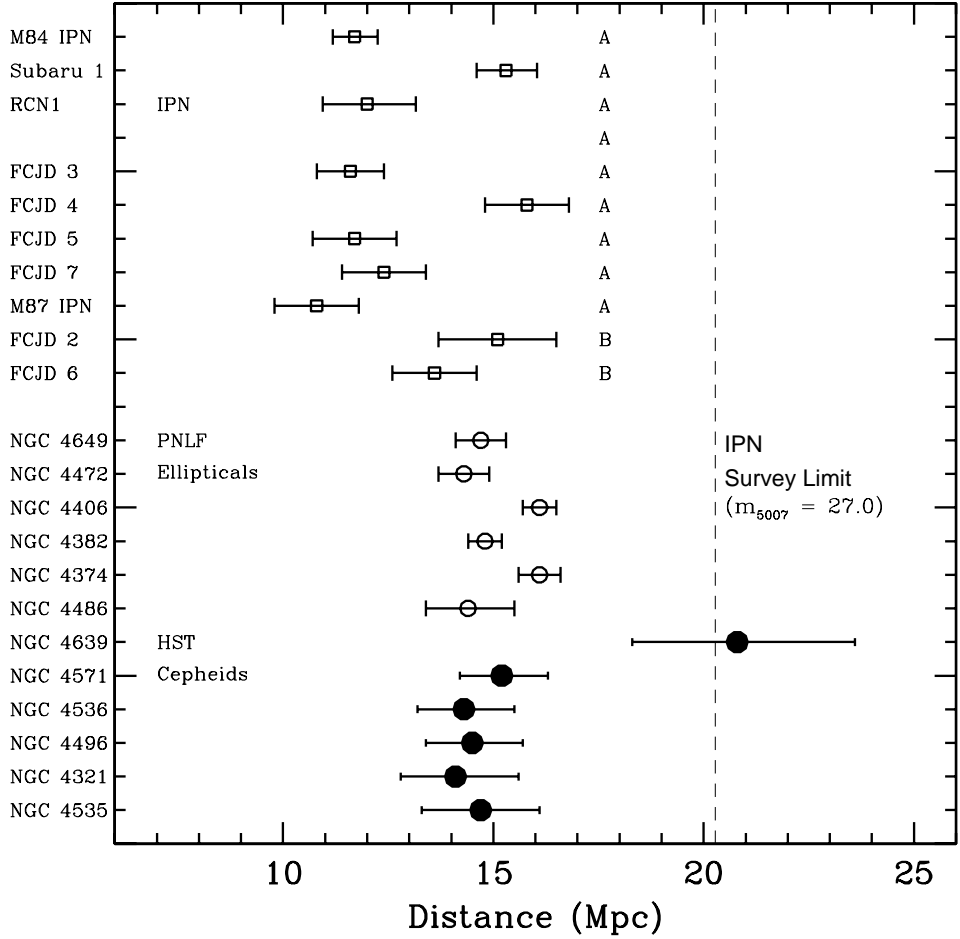}
\caption{The upper-limit distances to our and other IPNe fields compared
to PNLF distances of elliptical galaxies
and HST Cepheid distances of spirals.  As is clearly seen, some
of the IPN are up to 3 Mpc in front of the Virgo Cluster core.
The various subclumps of Virgo (A \& B) are denoted for each IPNe field.}
\end{figure}

\end{document}